\documentclass[preprint,12pt]{elsarticle}

\usepackage{amsmath}
\usepackage[english]{babel}
\usepackage{graphicx}

\begin{document}

\begin{frontmatter}

\title{Model of lateral diffusion in ultrathin layered films}

\author{Eugene B. Postnikov}
\ead{postnicov@gmail.com}
\address{Theoretical Physics Department, Kursk State University, Radishcheva st., 33 Kursk 305000, Russia}

\journal{arXiv}

\author{Igor M. Sokolov}
\ead{igor.sokolov@physik.hu-berlin.de}
\address{Institute of Physics, Humboldt-University at Berlin,
Newtonstr.15, 12489, Berlin, Germany}

\begin{abstract}
We consider the diffusion of markers in a layered
medium, with the lateral diffusion coefficient being the function
of hight. We show that the probability density of the lateral displacements 
follows one-dimensional Batchelor's equation with
time-dependent diffusion coefficient governed by the particles' redistribution in height.
For the film of a finite thickness the resulting mean squared displacement exhibits 
superdiffusion at short times and crosses over to normal diffusion at long times.
The approach is used for description of experimental results on inhomogeneous
molecular diffusion in thin liquid films
deposited on solid surfaces.

{\it \bf Highlights:}

\begin{itemize}
\item We show that vertical layering of liquid leads to superdiffusion behavior of walking markers
\item Finite film thickness cause the asymptotical transition from superdiffusion to normal diffusion
\item Conditional independence of orthogonal walks for 2D Fokker-Plank equation gives 1D Batchelor's
\end{itemize}

\end{abstract}


\end{frontmatter}

\section{Introduction}

Recent experimental research of diffusion and trapping of single particles of $\mu$- 
and nm-scales in confined liquid films \cite{Bevan2000,Schuster,Kihm,Banerjee2005,Huang,Sharma} show that 
this process presents non-trivial behavior considerably different from a usual 
Brownian diffusion in a bulk due to strong interaction of ultrathin fluid matrix  
with a base solid interface. The detailed single-molecule tracking 
detects the dependence of diffusivity on a distance of layer from the interface. 
It should be noted that a majority of evaluated experiments deals with the study of hydrodynamic interaction of colloid 
particles with a rigid wall that leads to change of both vertical and horizontal diffusivities \cite{Happel}. 
However, the very important parameter of such a system is a particle's size \cite{Bevan2000}. It has been found 
\cite{Carbajal,Banerjee2005,Huang,Sharma} that the experiments with micrometer-scale spherical particles diffusing
in a uniform fluid close to the wall quite satisfactory admit the two-dimensional depth-dependent 
anisotropic diffusion approach of \cite{Happel}.

At the same time, it was mentioned \cite{Kihm} that particle size change from $\mu$-m to nm scales leads to large deviations from hydrodynamic theory. 
Particularly, in the experimental work \cite{Kihm}  it was demonstrated that 
observed vertical diffusion coefficient is smaller. The discussion in \cite{Kihm} attributes this fact to the prevalence of intermolecular 
forces over hydrodynamical ones for such scales. From this point of view, very important experimental study was evaluated in the work \cite{Schuster}, 
where extremely small (molecular-size) markers were traced. It has bean detected there that the lateral diffusion is anomalous (superdiffusion) for short times and crosses over
to normal diffusion for long times. This phenomenon was discussed in a connection with the fact that fluids close to wetted interfaces are
know to form a layered structure. 

The motion of molecules in such layered structures can be described within the following random walk
picture. The layers close to the interface are practically frozen and
the lateral motion in these layers is slow. 
However, particles can diffuse (jump) vertically to layers,
where the  random walks are faster. Thus, the
transient diffusive behavior is based on a coexistence of these
vertical and lateral jumps. Note that the process of molecular diffusion 
in a wetting layer differs from the one of macroscopic particles near the wall
which can be described using the hydrodynamical picture \cite{Happel}, {\it vide infra}.

In what follows we adopt a coarse-grained description and consider the 
continuous model of layered diffusion: The particle performs diffusive 
motion in $y$-direction. The properties of this motion are independent on
the particle's lateral position due to translational symmetry of the fluid
layer in this direction, while the properties of diffusion in $x$-direction
depend on the particle's distance from the surface (i.e. on its $y$-coordinate).
Our model thus corresponds to a one of the layered system. 

The models of layered diffusion correspond 
to cases when the properties of motion in one ($y$) direction (normal to layers)
do not depend on the coordinate $x$, while the motion within layers (in $x$-direction) depends
on $y$. The properties of motions in $x$ and $y$-directions
differ, defining different models of layered diffusion:
Thus the combination of continuous drift in $y$ with random velocity model in $x$
defines Taylor's model of ``Diffusion by continuous motions'' \cite{Taylor}, the combination
of diffusion in $y$ with the same random velocity model in $x$ defines
a superdiffusive Matheron---De~Marsily model \cite{Matheron}, and combination of diffusion
in $y$ with transport with deterministic mean velocity proportional to $y$
defines the model of diffusion in a shear flow introduced by Novikov \cite{Novikov}. 
More general $y$-dependences of $x$-velocity define a class of models in \cite{BenNaim}.

The model of a layered system is not limited to hydrodynamical situations.
Thus, a similar model was recently used
for modeling of superdiffusion in a small world network
\cite{Postnikov}. Namely, the set of slow walks along a regular
background lattice and fast jumps along ``short links'' was replaced
by the set of simple diffusion equations with the hierarchy diffusion
coefficients depending on the distance between background nodes.

\section{The model}

A simple model described here adds an interesting example to the theory
of diffusion in layered systems and gives rise to an effective equation
of the Batchelor's type for lateral diffusion. Due to symmetry in the directions
parallel to the surface, it is enough to consider one lateral coordinate, which
is denoted $x$ in what follows. For this reason we consider 2D picture at first. 
Then we discuss the generalization on 3D case, which is quite straightforward.
Our model corresponds thus to a particle performing diffusion in $x$ and $y$-directions,
where the properties of the vertical diffusion (in $y$-direction) do
not depend on $x$. On a coarse-grained level (the steps of a random walk are not
resolved) this model is described by a pair of Langevin equations
\begin{eqnarray}
\dot{y} &=& \sqrt{2D_y} \xi_y(t) \label{motiony} \\
\dot{x} &=& \sqrt{2D_x(y)} \xi_x(t) \label{motionx}
\end{eqnarray}
with $D_y$ and $D_x(y)$ being the diffusion coefficients in $y$- and
in $x$-directions, and $\xi_y(t)$, $\xi_x(t)$ being Gaussian white
noises with zero mean and correlation property given by
$\langle\xi_\alpha(t) \xi_\beta (t') \rangle = \delta_{\alpha\beta}
\delta(t-t')$. Note that according to Eqs. (\ref{motiony}) and 
(\ref{motionx}) the motion in $y$ is independent on the motion
in $x$-direction, while the motion in $x$ does depend on the
actial $y$-coordinate through $D_x(y)$ and thus depends on the prehistory of $y$-motion. 
The random variables $x$ and $y$ are not independent.
 
The evolution of two-dimensional joint probability
density $p(x,y,t)$ for marker's position is then described by the
Fokker-Planck equation corresponding to the equation of anisotropic
diffusion:
\begin{equation}
\frac{\partial p}{\partial t}=D_x(y)\frac{\partial^2 p}{\partial x^2}+
D_y\frac{\partial^2 p}{\partial y^2}.
\label{init_eq}
\end{equation}
The Dirac delta function
$p(x,y,0)=\delta(x,y+0)$ is used as an initial condition.  We are
interested only in the time dependence of marginal distribution of
$x$, which is experimentally accessible \cite{Schuster}.
We note that although the situation with molecular tracer discussed here is different from the case of 
motion of a massive particle in a quiescent fluid, as discussed e.g. in \cite{Sancho82}, since
a molecular motion is never underdamped. However, the final equation is
exactly of the form obtained in this work, due to the fact that it is 
the only form corresponding to thermodynamical equilibrium (in the absence
of external forces) in a stationary state.

Although Eq.(\ref{init_eq}) itself does not have to possess a
solution which factorizes into a product of functions which depend
only on one of the two co-ordinates, the solution of our physical problem
given by Eqs.(\ref{motiony}), (\ref{motionx}) has to factorize: 
$p(x,y,t)=p(x,t)p(y,t)$, although $x$ and $y$ are not independent. 
The physical reason for this is as follows. 

First let us remark, that while the two arguments $x$ and $y$ of
$p(x,y,t)$ are the values of the corresponding random variables $x$
and $y$, its third argument, time, is not a random variable but essentially
an additional condition: $p(x,y,t)=p(x,y|t)$ is a conditional joint
probability of $x$ and $y$ provided their measurement took place at
time $t$. Note now that according to Eq.(\ref{motiony}) the motion in
$y$ direction is independent on $x$, and moreover the
initial condition for this motion in $y$ does not depend on
$x$. Therefore the conditional probability of $y$ given $x$ is
essentially a non-conditional one: $p(y|x,t) = p(y|t)$.  According to
Bayes theorem $p(x,y|t) = p(y|x,t)p(x|t) = p_y(y|t)p_x(x|t)$: the $x$
and $y$-displacements are {\it conditionally independent} \cite{Dawid1979}.
Rewriting it in our initial notation we get
\begin{equation}
p(x,y,t)=p_x(x,t)p_y(y,t).
\label{separation}
\end{equation}
Independently on whether this discussion is convincing or not, one can always prove that
the final solutions obtained by the separation ansatz are indeed the solutions of Eq.(\ref{init_eq}),
i.e. that the product of Eq.(\ref{sol1infy}) and Eq.(\ref{sol1infx}) indeed solves Eq.(\ref{init_eq})
in an unbounded system with $D_x(y)=ky$, and that the product of Eq.(\ref{sol1}) and of a symmetric Gaussian in $x$
with the dispersion given by Eq.(\ref{MSD}) solves it for a layer of a finite thickness, as they indeed do. 

Integrating both parts of Eq.(\ref{init_eq}) with respect to $x$ 
we obtain the equation for $p_y(y,t)$: 
\begin{equation}
\frac{\partial p_y}{\partial t}=D_y\frac{\partial^2 p_y}{\partial y^2}.
\label{eq_u_x}
\end{equation}

Inserting Eq.(\ref{separation}) into Eq.(\ref{init_eq}) and integrating 
over $y$ we obtain
\begin{equation}
\frac{\partial p_x}{\partial t}=\left[\int_0^H D(y) p_y(y,t) dy \right] 
\frac{\partial^2 p_x}{\partial x^2},
\label{eq_u_y}
\end{equation}
with $H$ being the film's thickness, 
i.e. the one-dimensional Batchelor's equation \cite{Batchelor}
\begin{equation}
\frac{\partial p_x}{\partial t}=D(t)\frac{\partial^2 p_x}{\partial x^2},
\label{Batchelor}
\end{equation}
with time-dependent diffusion coefficient
\begin{equation}
D(t)=\int_0^HD_x(y)p_y(y,t)dy.
\label{D}
\end{equation}
Note that this Batchelor's equation arises here not as the mean-field type
approximation, but as an exact equation for the corresponding marginal probability
density. Introducing a new variable $\tau=\int_0^tD(t)dt$ allows to reduce Eq. (\ref{Batchelor}) 
to the simple diffusion equation
\begin{equation}
\frac{\partial p_x}{\partial \tau}= \frac{\partial^2 p_x}{\partial x^2}.
\label{px}
\end{equation} 
Note that the change of variables from $t$ to $\tau$ is always possible within the model 
since the effective diffusion coefficient $D(t)$ given by Eq.(8) is always non-negative 
and therefore its integral over time is a monotonically growing (and therefore invertible) function.
From this it follows
that the corresponding probability density is Gaussian with a 
mean square displacement (MSD) $<x^2>=2\tau$.

In a 3D system uniform in $(x,z)$ coordinates with $D_z(y)=D_x(y)$ still depending on the vertical 
$y$ position, the system of Eqs.~(\ref{motiony})--(\ref{motionx}) is supplemented by 
\begin{equation}
\dot{z}=\sqrt{2D_x(y)} \xi_z(t). \label{motionz}
\end{equation}

Thus, from the Eqs. (\ref{motiony})--(\ref{motionx}), (\ref{motionz}) and the symmetry in the directions
parallel to the surface, it follows that the three-dimensional joint probability
density $p(x,y,z,t)$ satisfies the equation:
\begin{equation}
\frac{\partial p}{\partial t}=D_x(y)\left[\frac{\partial^2 p}{\partial x^2}+\frac{\partial^2 p}{\partial z^2}\right]+
D_y\frac{\partial^2 p}{\partial y^2}.
\label{init_eq3D}
\end{equation}

As before, all displacements are conditionally independent 
and 
the three-dimensional joint probability
density will be factorized as 
\begin{equation}
p(x,y,z,t)=p_x(x,t)p_y(y,t)p_z(z,t).
\label{3Dfactorization}
\end{equation}

The equation for $p_z(z,t)$ has exactly the form of the Eq.~(\ref{px}).

Finally, since the motions in both lateral directions are independent, the resulting MSD is
\begin{equation}
<r^2>=<x^2>+<z^2>=2<x^2>=4\tau.
\label{r2}
\end{equation} 
In other words, it is simply two times larger then in 2D case, and the definition of $\tau$ follows from the Eq.~(\ref{D}) as above.

\section{Solutions}

\subsection{Superdiffusion in a thick film}

At first, we consider an idealized situation: the half-plane corresponding 
to the film of infinite thickness $H=\infty$. In this case the solution of 
Eq. (\ref{eq_u_x}) is the Gaussian
\begin{equation}
p_y(y,t)=\frac{1}{\sqrt{D_y\pi t}} \exp \left(-\frac{y^2}{4D_yt}\right).
\label{sol1infy}
\end{equation}
Thereafter, one needs to determine the vertical dependence of the
horizontal component of the diffusion coefficient. As it will be
shown below, the experimental data are well-fitted
by a linear dependence $D_x(y)=ky$. Under this condition, the substation
(\ref{sol1infy}) into (\ref{D}) gives
\begin{equation}
D(t)=2k\sqrt{D_yt/\pi}.
\label{Dtinf}
\end{equation}
Therefore, the averaged lateral mean-square displacement corresponds to 
the superdiffusion:
\begin{equation}
\left<x^2\right>=\frac{8k}{3}\sqrt{\frac{D_y}{\pi}}t^{\frac{3}{2}},
\label{MSDinfthick}
\end{equation}
and the corresponding distribution of lateral positions is given by a
Gaussian
\begin{equation}
p_x(x,t)=\frac{\exp \left(-\frac{x^2}{\sqrt{(4/9\pi)k^2D_y^3t^3}} \right)}{\sqrt[4]{(64/9)\pi k^2D_y^3t^3}}.
\label{sol1infx}
\end{equation} 

 As it was discussed above, in 3D case additional multiplier $p_z(z,t)$ in the Eq.~(\ref{3Dfactorization}) 
will have exactly the same form as (\ref{sol1infx}) with replacing $x \to z$, and MSD $<r^2>$ is given by 
twice the expression (\ref{MSDinfthick}).

\subsection{Film of finite thickness}

Now let us consider the realistic situation, when the thickness $H$ is
finite. In such a case the equation (\ref{eq_u_x}) with impenetrable
boundaries has a standard solution in the form of the eigenfunction expansion:
\begin{equation}
p_y(y,t)=\frac{1}{H}+\frac{2}{H}\sum_{n=1}^{\infty}
e^{-\frac{D_y\pi^2n^2}{H^2}t}\cos\left(\frac{\pi n}{H}y\right).
\label{sol1}
\end{equation}
Under the same condition $D_x(y)=ky$, the time dependence of the
averaged coefficient of lateral diffusion is represented by the
 series
\begin{equation}
D(t)=\frac{kH}{2}\left(
1-\frac{8}{\pi^2}\sum_{m=0}^{\infty}\frac{e^{-\frac{D_y\pi^2(2m+1)^2}{H^2}t}}
{(2m+1)^2}
\right).
\label{Dt}
\end{equation}
From Eq. (\ref{Dt}) it follows that the effective diffusion
coefficient is initially equal to zero due to sticking of a
marker to the solid substrate, monotonically grows with time and then saturates. 
Its final value $D(\infty)=kH/2$ is achieved when the markers redistribute homogeneously
over the whole liquid layer. 

Integrating Eq. (\ref{Dt}) over time we
get the desired expression for a mean-square-displacement:
\begin{equation}
<\hspace{-1mm}x^2\hspace{-1mm}>=kH\left(
t+\frac{8H^2}{D_y\pi^4}\sum_{m=0}^{\infty}\frac{e^{-\frac{D_y\pi^2(2m+1)^2}{H^2}t}-1}
{(2m+1)^4}
\right).
\label{MSD}
\end{equation}

Analyzing Eq. (\ref{MSD}), one can see that non-linear growth at a small
time changes to the asymptotic linear one  corresponding to a normal diffusion:
\begin{equation} 
<x^2\hspace{-1mm}>_{\infty}=-\frac{kH^3}{12D_y}+kHt.
\label{x2norm}
\end{equation}

\begin{figure}
\includegraphics[width=\columnwidth]{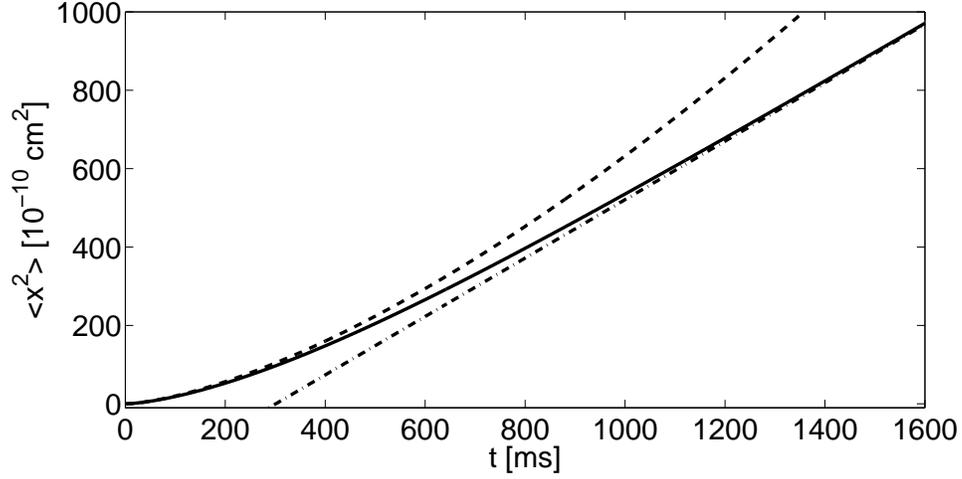}
\caption{The time dependence of mean-square-displacement (solid line)
for the parameters: $H=17 \; \mathrm{nm}$, $k=0.4384 \; \mathrm{cm^2/(s\cdot
nm)}$, $D_y=8\cdot 10^{-13} \; \mathrm{cm^2/s}$. Two additional referent
lines correspond to the superdiffusion $<x^2>=0.02t^{3/2}
10^{-10} \; \mathrm{cm^2}$ (dashed line) and the asymptotic normal
diffusion $<x^2>_{\infty}$. (dash-dotted line)}
\label{MSDfig}
\end{figure}

Fig.~\ref{MSDfig} shows the typical behavior of the mean squared displacement
as a function of time. It clearly shows the non-stationary character of
diffusion in ultrathin liquid film. During the first stage the
motion is superdiffusive: the curve is well fitted
by power-law function with the exponent $3/2$. 
The microscopic physical background of this can be explained
by jumps of markers into higher layers and fast transport there.
The further saturation of layers by markers leads to decrease of the
relative contribution of processes ``jump to a higher layer and fast
diffusion there''. Correspondingly, the growth of the
mean squared displacement decelerates. Finally, the uniform
redistribution of markers changes a kind of transport to a normal
diffusion.

All corollaries concerning simple 3D generalizations are the same as in the previous subsection.

\section{Comparison with experiment}

\begin{figure}
\includegraphics[width=\columnwidth,height=0.6\columnwidth]{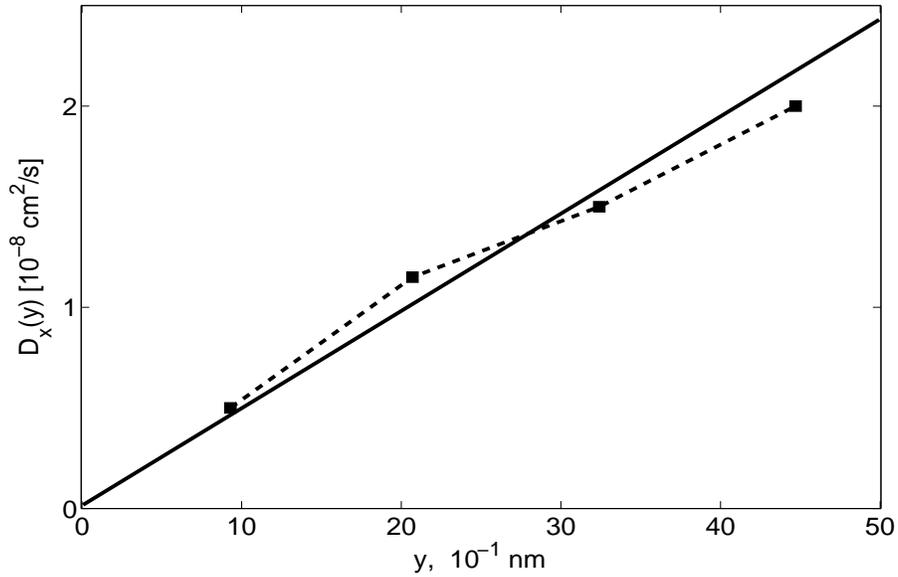}
\caption{The linear fit of experimental data of Ref. \cite{Schuster} (Fig. 6b there; black squares corresponds to the same symbols in the cited source), which
confirms the chosen dependence $D_x(y)=ky$ and allows to determine the
tangent $k=0.4384 \; \mathrm{cm^2/(s\cdot nm)}$.}
\label{fit}
\end{figure}

To show that the model presented does make sense, the comparison with direct measurements
of inhomogeneous molecular diffusion in thin liquid film deposited on
solid surface \cite{Schuster} is carried out. Let us first show, that
the assumption of the linear dependence of the microscopic lateral diffusion
coefficient on the distance from solid substrate holds. The least-square linear fit 
of  the experimental data for lateral diffusion coefficients taken from \cite{Schuster} 
is presented Fig.~\ref{fit}.  This fit  justifies the assumption the approximate law for 
$D_x(y)$ with a rather good accuracy.

\begin{figure}[t]
\includegraphics[width=\columnwidth]{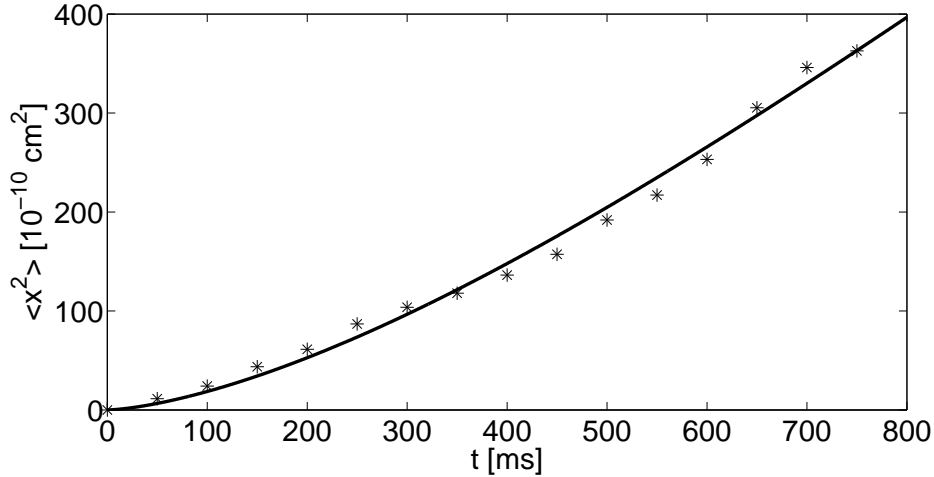}
\caption{The comparison of calculated curve of a
mean-square-displacement  (solid line) with experimental data (asterisks) provided by authors of
\cite{Schuster}  for short observation time. Parameters of the film: $H=17 \; \mathrm{nm}$, $k=0.4384
\; \mathrm{cm^2/(s\cdot nm)}$; vertical diffusion coefficients
$D_y=8\cdot 10^{-13} \; \mathrm{cm^2/s}$  (or $8\cdot 10^{-13}/\sqrt{2} \; \mathrm{cm^2/s}$) for 3D case.}
\label{MSDexp}
\end{figure}

Using the value of the slope obtained, the calculation according to the
formula (\ref{MSD}) are completed for the experimental film thickness
(note that the series (\ref{MSD}) has a relatively fast convergence, e.g. the maximal relative 
difference between result presented by truncations at 11th and 10th terms is 2\% and at between 
truncated 21th and 20th terms is 0.2\%). Note, that the vertical component of diffusion
coefficient is unknown from the direct measurements. Therefore, it is
estimated in such a way that the resulting curve fits the experimental
points, see Fig.~\ref{MSDexp}. These experimental points represented as stars in Fig.~\ref{MSDexp}.
 Note also that if one need to take into account 3D picture with two-dimensional lateral layers, the curves keep the same shape, but the adjustable parameter would be simple $\sqrt{2}$ times smaller. This follows from the Eqs.~(\ref{D}), (\ref{r2}).

The experimental points correspond to a single trajectory from the set of the overall 4 trajectories 
kindly provided by authors of Ref. \cite{Schuster}. 
For all of them the behavior at short times (governed by practically deterministic process
of the departure from the surface) is practically the same.
For longer observation times, the individual trajectories show various fluctuation features 
like large excursions and returns, and the averaging over our very restricted sample does not produce
better results than taking an individual one. We use the trajectory, MSD of which corresponds 
to most regular curve over the time interval considered.

The considered time range allows to cover two specific regimes: 
i) During the interval from 0 to approximately 400~ms, the process is superdiffusive 
MSD ($<x^2>\sim t^{3/2}$). It coincides with the analytically exact superdiffusion 
obtained for semi-infinite region. The natural reason for this is that at short times a walker can 
not approach the upper boundary of film. Thus there is no sufficient difference 
between solutions as well as for physical experimental picture.  ii) The transient 
process from superdiffusion to normal diffusion starting from time approximately 400~ms. 
Within this subinterval, the MSD curve can not be represented as a power-law function, 
compare with the Fig.~\ref{MSDfig}, however the series solution (\ref{MSD}) adequately 
reproduces experimental data. 

\begin{figure}[t]
\includegraphics[width=\columnwidth]{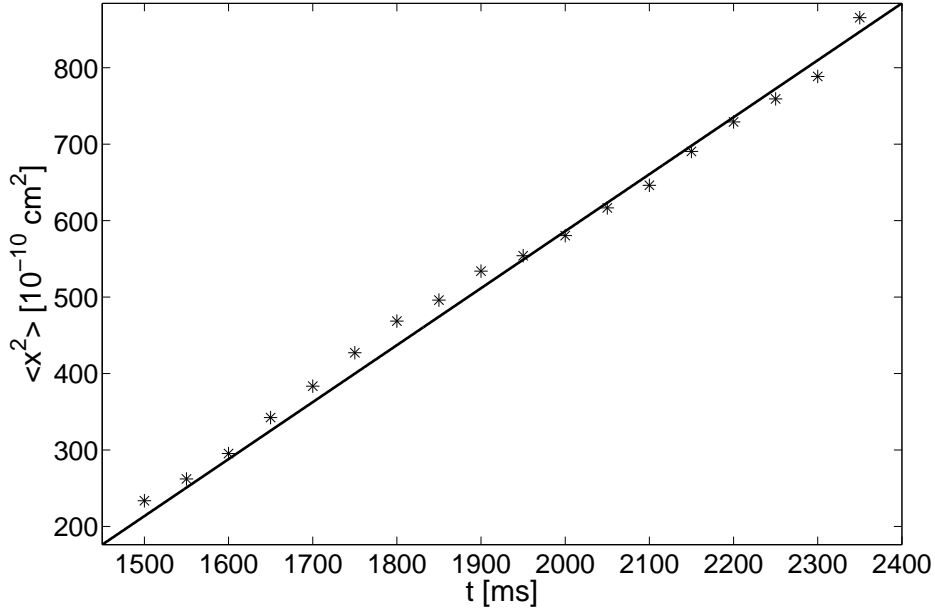}
\caption{ The comparison of calculated curve of a
mean-square-displacement ($<x^2\hspace{-1mm}>_{\infty}\sim kHt$, see the Eq.~(\ref{x2norm}). The constant vertical shift does not coincide with the analytical expression due to certain irregular markers' jumps and stops during the experiment) with experimental data (asterisks) provided by authors of
\cite{Schuster}  for long observation time. Parameters of the film: $H=17 \; \mathrm{nm}$, $k=0.4384
\; \mathrm{cm^2/(s\cdot nm)}$; vertical diffusion coefficient
$D_y=8\cdot 10^{-13} \; \mathrm{cm^2/s}$ (or $8\cdot 10^{-13}/\sqrt{2} \; \mathrm{cm^2/s}$ for 3D case).}
\label{MSDnormdiff}
\end{figure}

The asymptotic large-time MSD corresponding to the normal diffusion, 
see Fig.~\ref{MSDfig}  for simulated curve and Fig.~\ref{MSDnormdiff} confirming this conclusion by the experimental data. See also discussion of long observation times in \cite{Schuster}.

 Thus, that anomalous diffusive
behavior in lateral direction is caused by very weak vertical
diffusion, i.e. rare jumps between layers: their rate is five orders of magnitude
smaller then the lateral one. 

Note also that our model deals with averaged
motion. Correspondingly, it does not take into account fluctuations
along individual trajectories, which are significant for large times of observation
mirrored in growing fluctuations in $D$ \cite{Schuster}. 
Since the total number of trajectories obtained in the experiment is limited, and these trajectories show considerable differences between different runs, as they should. Fitting the results of a theory relying on ensemble-averaged quantities to this data is therefore connected with large statistical uncertainties, and one can mostly speak about trends, which are correctly captured by our theoretical description. The more convincing proof can be only given by performing more experiments in longer runs.

\section{Additional discussion}

One of our basic simplifications is based on the assumption of a
constant vertical diffusion coefficient coexisting with
height-dependent lateral one.  However, detailed studies
of a near-wall motion of mesoscopic particles show that both 
components of the diffusion coefficient should depend on the
distance from a wall, see the theoretical derivation and
discussion in \cite{Happel}.
In this case the explicit $y$-dependence of the vertical diffusion coefficient
should be taken into account, and
Eq.~(\ref{init_eq}) should be replaced by a more general one
\begin{equation}
\frac{\partial p}{\partial t}=D_x(y)\frac{\partial^2 p}{\partial x^2}+
\frac{\partial}{\partial y}D_y(y)\frac{\partial p}{\partial y}.
\label{init_eq_cd}
\end{equation}
(If starting on the Langevin level, the kinetic (Hanggi-Klimontovich) interpretation
of the Langevin equation for vertical motion should be used when assuming that the
equilibrium distribution of the molecules in $y$-direction is homogeneous \cite{Sokolov2010}).
Although Eq.~(\ref{init_eq_cd}) is more complex than our initial Eq.~(\ref{init_eq}), 
the assumption of independence of motion in $y$ on the lateral position and thus the
conditional independence and the factorization of $x$- and $y$-motions still hold.

Now, let us subdivide the vertical diffusion coefficient in (\ref{init_eq_cd}) into 
a constant and a variable parts: $D_y(y)=D_y^{(c)}+D_y^{(v)}(y)$. 
Then the Eq.~(\ref{eq_u_x}) takes the form:
\begin{equation}
\frac{\partial p_y}{\partial t}=
\left(D_y^{(c)}+D_y^{(v)}(v)\right)\frac{\partial^2 p_y}{\partial y^2}+
\frac{\partial D_y^{(v)}}{\partial y}\frac{\partial p_y}{\partial y}
\label{eq_u_x_mod}
\end{equation}
Supposing that $D_y^{(v)}<<D_y^{(c)}$ we can neglect the second term in 
brackets in the first term in the right hand side of this equation (in a
factor in front of the second derivative) and conclude that the coordinate
dependence of $D_y$ mainly leads to the emergence of a drift
term in $y$-direction. The last one actually changes the motion from a 
pure Brownian one to a combination of diffusion and regular drift. These results in a
change of the effective MSD-exponent in $y$ from 1.5 closer to 2 during 
the first stage but does not influence the late stage of the process: after a uniform
redistribution of walkers over a bulk the averaged lateral MSD will
correspond to the normal diffusion. The numerical solution of
(\ref{eq_u_x_mod}) with $D_y^{(v)}(y)=0.05\div0.2D_y^{(c)}y$ shows that initial superdiffusion in this
case is sufficiently faster than the experimentally observed one, 
and moreover that the transient to a normal diffusion is shorter. 
In fact the resulting process comprises short-time regular ballistic drift 
($\left<x^2\right>\sim t^2$) leading to the uniform filling of the layer with 
succeeding normal lateral diffusion. 

Thus, the
assumption of uniform small vertical diffusion coefficient can be considered as 
adequate for the description of the observed pattern of  lateral diffusion  for extremely small tracers, confirming the supposition made in \cite{Kihm}. The additional
confirmation of this assumption can be found in the work \cite{Schuster2003},
where the interpretation of experiments is provided. Authors argued
that the microscopic molecular picture of media corresponds to the
layered structure with small probability of jumps of molecules
(nanometer scale objects) between adjacent layers.  Our model
discussion confirms these conclusions. 

\section{Summary}
We considered lateral diffusion of a particle in a layered ultrathin fluid medium and obtained the
solutions for the lateral mean-square-displacement. The results demonstrate the transition 
from superdiffusion at short times to asymptotically normal diffusion, as 
is confirmed by the experimental findings.

From the theoretical point of view, our work discusses an interesting case when the Batchelor's equation in 1D
appears to be exact, emerging from the full two-dimensional Fokker-Plank equation  
due to conditional independence of motions along both coordinates. Note that these conditionally independent
motions are still connected via the common parameter: the time.

\section*{Acknowledgments}
We thank Prof.~C.~von~Borzcyskowski and Dr.~J.~Schuster for provided
experimental data and valuable discussions.

\end{document}